\documentstyle[fleqn,11pt]{article}
\newcommand{\beq}{\begin{equation}}
\newcommand{\eeq}{\end{equation}}
\title{\bf Detection of brown dwarfs by the micro-lensing of
unresolved stars}

\author{Paul Baillon\thanks{CERN, 1211 Gen\`eve 23, Suisse} \and Alain
Bouquet\thanks{Laboratoire de Physique Th\'eorique et Hautes
Energies, Universit\'es Paris 6 and Paris 7, Unit\'e associ\'ee au
CNRS (UA 280), 2 Place Jussieu, 75251 Paris Cedex 05 FRANCE} \and
Yannick Giraud-H\'eraud\thanks{Laboratoire de Physique
Corpusculaire, Coll\`ege de France, Laboratoire associ\'e au
CNRS-IN2P3 (LA~41), 75231 Paris Cedex 05 FRANCE}
\and Jean Kaplan$^\dagger$}

\date{}

\begin{document}
\vspace{2cm}
\maketitle
\vspace{2cm}

\begin{abstract}
\vspace{1cm}
{\large The presence of brown dwarfs in the dark galactic halo could be
detected through their gravitational lensing effect and
experiments under way monitor about one million stars to observe a
few lensing events per year. We show that if the photon flux from a
galaxy is measured with a good precision, it is not necessary to
resolve the stars and besides more events could be observed.}
\end{abstract}
\vspace{1cm}

\vspace{2cm}
\noindent PAR-LPTHE 92 39\\
\noindent LPC 92 17

\vspace{1cm}
\noindent {\em Submitted to Astronomy \& Astrophysics}
\newpage

\section{A modest proposal}

Many observations suggest that galaxies are embedded in massive
spheroidal dark halos (Faber \& Gallagher 1979; Kormandy \& Knapp 1987;
Trimble 1987; Turner 1990)
which could be made of compact baryonic objects such as jupiters or
brown dwarfs (Carr \& Bond 1984; Rees 1986). Paczynski showed that such dark
objects could be detected through gravitational lensing~: the light
of a star is amplified when a brown dwarf gets in the line of sight
(Paczy\'nski 1986). As the brown dwarf moves relative to the line
of sight, this amplification varies in a characteristic way with
time. The main problem is the low probability of such a
micro-lensing event~: about 10$^6$ stars must be monitored for one
year to observe one 30 \% luminosity increase. Nevertheless,
several experiments are under way (Alcock et al. 1989; Moniez 1990;
Ansari 1992).

We suggest that, instead of monitoring individual stars, one could
detect the flux amplification due to the lensing of one {\em
unresolved} star in a dense star field, if the global luminosity of
the field can be measured with sufficient accuracy. Then {\em
every} star in the field is a candidate for a micro-lensing, not
only the few resolved stars, and therefore the event rate is
potentially much larger. Of course, not all micro-lensing events
can be detected. Since the photon flux coming from a star is a
small fraction of the flux coming from the field, only the lensing
of bright stars can be detected (but they are rare), unless the
amplification is very strong (but this is a rare event also because
it requires a close alignment of the brown dwarf with the star).
There are two promising targets, the central bar of the Large
Magellanic Cloud (LMC) and the Andromeda galaxy M31. Our proposal
can be implemented with photon counters or with CCD cameras (better
angular resolution but less precision on the flux). We report here
the result of both studies which conclude to a higher efficiency
than the standard procedure. Details of the method and full
computations will be described in forthcoming papers. We first
estimate the photon flux received on a pixel, the required
amplification needed to detect an increase and the corresponding
event rates (after recalling a few basic formulas for
micro-lensing).

\section{Fluxes and amplification}

To evaluate the required amplification, we need to know the number
of photons received on one pixel of the detector during an
exposure, from the lensed star and from the host galaxy. It is easy
to evaluate the photon flux $F_{star}$ received from a star of
given magnitude, if we approximate it by a black body, at a
temperature given by its color index $B-V$. The flux received on
the pixel from the galaxy is just the sum $F_{gal} = \sum_{stars}
F_{star}$ of the fluxes from the stars which lie in the field of
view of the pixel. This sum is given by the integral over the
luminosity function of the target galaxy, i.e. the density of stars
as a function of their magnitude and spectral class. This function
is not well known~: it is measured in limited areas, for bright
stars only, and the only spectral information we have is a separation between
main sequence stars and red giants (Hardy et al. 1984; Ardeberg et al. 1985 ;
Hodge et al. 1988).
For lack of a better information, we use the quoted data, assuming all
red giants to be in the K class, and we completed them at the faint end
by the
luminosity function of the solar neighborhood (Allen 1973), with an
 empirical relative normalisation.
The resulting luminosity function is then normalised to
the measured surface brightness of the galaxy in the area covered
by the pixel (de Vaucouleurs \& Freeman 1972; Kent 1983; Walterbos
 \& Kennicutt 1987). We checked that the expected rate of micro-lensing
events is not very sensitive to the chosen luminosity function (see Figure 4).
The total flux
received on the pixel is the sum of this flux $F_{gal}$ and of the
flux $F_{back}$ from the night sky background (Allen 1973). We
take into account the absorption, and the quantum efficiency of the
detector, which reduce the fluxes to a fraction of the photons
arriving on top of the atmosphere. The total number of photons
recorded during an exposure time $t_{exp}$ on a pixel with a
collecting area $S$ is $N_{tot} = (F_{gal} + F_{back}) \,S
\,t_{exp}$~.

This number actually is a random variable, with mean value
$N_{tot}$ and standard deviation $\sigma$~:
\beq
\sigma \,=\, \sqrt{N_{tot}} \,=\, \left[ \left( \sum_{stars}
F_{star} + F_{back} \right) \,S \,t_{exp} \right]^{1/2}
\label{sigma}
\eeq
Statistical fluctuations therefore limit the relative accuracy of
the photometry to be worse than $\sigma/N_{tot} = 1/\sigma$. To get
a feeling of the numbers involved, we consider two reference detectors, a CCD
camera and a photomultiplier array, the characteristics of which are given in
Table 1 below. The
statistical accuracy is about 2\% for the CCD camera pointing to an area of
$\mu=$21 mag/arcsec$^2$ for an exposure time $t_{exp}=1/4$ h, and
about 10$^{-4}$ for the photomultiplier array in the same
conditions. Longer exposure times, or larger collecting areas,
improve the situation. However, there are unavoidable sources of
spurious fluctuations, such as a luminous object passing through
the field of view, a change in the background light or in the
transparence of the atmosphere\ldots Such effects limit in practice the
photometric
accuracy, and will probably require two identical telescopes
operating in coincidence (as in one proposed Th\'emistocle
upgrade). Another source of unwanted fluctuations comes from the
limited pointing accuracy of the telescope~: the number of stars on
a pixel is also a random variable, and it fluctuates from one pixel
to another. An imperfect pointing of the telescope then makes the
matching of pixels from one image to the next more difficult, and
this induces spurious luminosity fluctuations. Work is now in
progress to assess and overcome the effects of these pointing
uncertainties, but they are neglected in this paper.

To claim a positive observation of a micro-lensing, we require a
minimal amplification $A_{min}$ such that the increase $(A_{min} -
1) F_{star} \,S\, t_{exp}$ in the number of photons is $Q$ times
($Q = 3, 5, 10 \ldots $) the standard deviation $\sigma =
\sqrt{N_{tot}}$. Then~:
\beq
A_{min} = 1 + Q \frac{\sqrt{(F_{gal} + F_{back}) \,S \,t_{exp}}}{
F_{star}\, S\, t_{exp}}
\label{amin}
\eeq
When the pixel aperture is very small, around 1 arcsec$^2$ as on
CCD cameras, Equation \ref{amin} must be corrected. The mean number
of stars on a pixel is small, and one star may dominate the
luminosity of the pixel. When a bright star sits on a pixel, the
flux $F_{star}$ received from this star alone can be larger than
the mean flux $<F_{gal}>$ received on neighboring pixels, and we
use the formula $F_{gal} \simeq F_{star} + <F_{gal}>$. Seeing is
another complication for small apertures~: only a fraction $f$ of
the flux of a star reaches the pixel, and we must replace
$F_{star}$ by $fF_{star}$ . The remaining light is spilled over
neighboring pixels, but this is usually compensated by the spilled
light from neighboring pixels, unless the light distribution of the
target galaxy is very irregular. Hence, Equation \ref{amin} is
modified as~:
\beq
A_{min} = 1 + Q \frac{\sqrt{(fF_{star} + <F_{gal}> + F_{back}) S
t_{exp}}}{f F_{star} S t_{exp}}
\label{amin2}
\eeq
Seeing effects spread the light of a star over several pixels and
increase the amplification needed to detect a micro-lensing. On the
other hand, this spread also reduces the spurious fluctuations due
to bad pixel matching.

Contrary to our naive intuition, it turns out that amplifications
are not very large for detectable lensing events. Figure 1 shows
that amplifications are seldom much greater than 10, for both
target galaxies. We did find some large amplifications
($A_{min}>100$) in our Monte-Carlo simulations, but they are very
rare. It appears to be more efficient to weakly amplify a few
bright stars than to strongly amplify many faint stars, and this
can be traced back to the slope of the luminosity functions
(Allen 1973; Hardy et al. 1984; Hodge et al. 1988).
Indeed, the lensed stars (for
detectable events) are mostly main sequence stars in the LMC, and
red giants in M31 (see Figure 2).

\section{Basics of micro-lensing}

When a massive object comes at a distance $R$ to the line of sight
of a star, the star light is amplified by a factor $A(R)$~:
\beq
A(R) = \frac {R^2+2R_E^2}{R\sqrt{R^2+4R_E^2}}
\label{ampli}
\eeq
where the Einstein radius $R_E$ is defined by~:
\beq
R_E^2 = \frac{4 G M_{bd}}{c^2} \frac{D (D_{star}-D)}{D_{star}}
\eeq
for a deflector of mass $M_{bd}$ at a distance $D$, and a star at
distance $D_{star}$ from the observer (Paczy\'nski 1986). Note
that, for $R \ll R_E$ the amplification $A \simeq R_E/R$. Brown
dwarf masses range (De R\'ujula et al. 1992) from $10^{-7}\, M_{\odot}$
(evaporation limit) up to $10^{-1}\, M_{\odot}$ (hydrogen burning
limit). The Einstein radius sets the scale of the lensing~: for
instance, the amplification stays larger than $A$ for a time
$t_{event}\simeq R_E/A V_{\perp}$ if the brown dwarf has a
transverse velocity $V_{\perp}$. Numerically~:
\beq
t_{event} \simeq 10^5\,{\rm s}\, \frac{1}{A} \, \frac{200 \, {\rm
km/s}}{V_{\perp}}\,\left( \frac{M_{bd}}{10^{-4}\,M_{\odot}}
\frac{D}{25\,{\rm kpc}} \frac{D_{star}-D}{D_{star}} \right) ^{1/2}
\eeq
The event rate $\Gamma_{star}$ (number of micro-lensing events that
can be detected per unit time) for a given star is simply the
number of brown dwarfs which come close enough to the line of sight
to yield a mean amplification, during the
exposure time, above some threshold $A_{min}$. This
number depends on the required amplification $A_{min}$, on the
brown dwarf number density in the direction of the star, and on the
brown dwarf mean transverse velocity $V_{\perp}$. The total number
of micro-lensing events detected during the observation time
$t_{obs}$ then is the sum over all stars of the number of events
for each star~:
\beq
N_{events} = t_{obs} \sum_{stars} \Gamma_{star}
\eeq
We assume that our Galaxy is embedded in a massive dark halo made
of brown dwarfs, described by an approximate isothermal
distribution with a core radius $a$. We took all brown
dwarfs to be of equal mass $M_{bd}$. Of course, we do not expect this
to be true, but i) the actual mass distribution is totally
unknown, and ii) this allows to probe the sensitivity of different
experimental conditions for various brown dwarf masses. The brown dwarf number
density
$n(r)$ is~:
\beq
n(r) = \frac{\rho_{\odot}}{M_{bd}} \frac{r_{\odot}^2+a^2}{r^2+a^2}
\eeq
It is cut-off at a maximal radius $r_{max}$ which is unknown, and
we shall assume that it extends up to 100 kpc in numerical
estimations. The halo mass density $\rho_{\odot}$ in the solar
neighborhood is estimated to be 0.008 $M_{\odot}/{\rm pc^3}$
(Flores 1988). The distance $r_{\odot}$ between the Sun and the
galactic center is 8.5 kpc, and estimates of the core radius range
from $a = 2$ kpc (Bahcall \& Soneira 1980) to $a = 8$ kpc (Caldwell
\& Ostriker 1981),
and we shall take $a = 5$ kpc for definiteness. We also assume that
brown dwarfs have a Maxwellian velocity distribution, with a
uniform velocity dispersion of 270 km/s.

Equation \ref{ampli} is only valid for point sources, and should be
modified for extended sources such as stars. Actually, the shape of
$A(R)$ is not very different from the point-like case, except that
there is an upper limit $A_{max}$ on the amplification of an
extended source by micro-lensing. This limit is reached in the case
of perfect alignment (where the amplification of a point-like star
would be infinite). A surface element of the star at a distance $y$
from the center is then amplified by a factor $A(y)$~:
\beq
A(y) = \frac{(yD/D_{star})^2 +
2R_E^2}{yD/D_{star}\sqrt{(yD/D_{star})^2 + 4R_E^2}}
\eeq
and~:
\beq
A_{max} = \frac{1}{\pi R_{star}^{2}} \int_{0}^{R_{star}} A(y) 2 \pi
y dy = \left[ 1 + \left( 2 \frac{R_E(D) / D}{R_{star}/D_{star}}
\right) ^2 \right] ^{1/2}
\label{amax}
\eeq
For a given star and a given brown dwarf mass, this leads to an
upper bound $D_{max}$ on the distance $D$ of a brown dwarf, which
becomes small for large amplifications or large stellar radii (the
radius of a given star is related to its spectral class and
magnitude (Allen 1973)). This effect reduces the number of
detectable lensings of red giants, or the lensing efficiency of
lighter brown dwarfs. For instance, Figure 3b shows that light
brown dwarfs of the halo of M31 do not lead to detectable lensings.

\section{Number of events}

We use two methods to compute the expected number of micro-lensing
events~: a Monte-Carlo simulation, and a semi-analytic method. The
results obtained are very similar, and allow a cross-check of the
accuracy of both computations.

Our Monte-Carlo procedure is the following. We choose a
brown dwarf of fixed mass with random position in the galactic halo
(and also in the halo of M31 when
this galaxy is the target) according to the
brown dwarf density distribution $n(r)$, and with random velocity
according to a Maxwellian distribution . Around the line of sight between the
Earth and this brown dwarf we define a `useful angular area'
characterized by the Einstein angle $R_E(D)/D$ and the extrapolated
angular trajectory of the brown dwarf during the year. We then pick
up a uniformly distributed star in the projection of this area on
the plane of the target galaxy. In order to keep the computation
time reasonable, we affect to each event a weight reflecting the
density, magnitude and spectral distribution of stars. This weight
is eventually normalised to the local surface brightness.  In a
second step, we follow the trajectory of the brown dwarf for one
year and check wether it gets close enough to a star to lead to a
detectable amplification of the starlight. When large
amplifications are required, micro-lensing events are short, their
duration becomes comparable to the exposure time $t_{exp}$ and the
amplification changes during the exposure. We thus compute the mean
amplification $\bar{A}$ during the exposure and require $\bar{A} >
A_{min}$. We store the corresponding event for further analysis,
and repeat the whole procedure, a few million times for each value
of the unknown parameters (the mass distribution of brown dwarfs,
the exposure time, the detector characteristics, etc.).

In the semi-analytic method, we again compute for a given star the
mean amplification $\bar{A}$ during the exposure and require
$\bar{A} > A_{min}$. For each star, this defines a volume $\Delta$
around the line of sight, in which a brown dwarf must lie at time
$t = 0$ (the middle of the exposure) to yield an observable effect.
The number of micro-lensing events which can be seen during a given
exposure is nothing but the number of brown dwarfs which lie in the
volume $\Delta$ of each star, summed over all stars in the field of
view of the pixel. This number of brown dwarfs is just the integral
over $\Delta$ of the number density $n$, and the sum over stars is
the integral over the luminosity function (normalised to the
surface brightness of the galaxy). The amplification can remain
above threshold for several exposures, and to avoid multiple
counting, one must only count for each exposure brown dwarfs which
were not yet inside the volume during the preceding exposure. These
brown dwarfs, efficient for the first time, lie in a smaller volume
$\Delta '$, obtained by
substracting from volume $\Delta$ the volume $\Delta$ shifted by
$\vec{ V_{\perp}}(t_{inter}+t_{exp})$ where $t_{inter}$ is the time
interval between the two exposures, and $\vec{ V_{\perp}}$ the
transverse velocity of the brown dwarf. The total number of
micro-lensing events during the whole observation is the number of
new events per exposure, times the number of exposures $N_{exp}$
done during the total observation time $t_{obs}$, i.e.~:
\beq
N_{events} = N_{exp} \sum_{stars} \int_{\Delta '} n(r,V_{\perp})
{\rm d}^3 r {\rm d}^2 V_{\perp}
\eeq
We recover the usual formula (Griest 1991) when the exposure time
$t_{exp}$ is much shorter than the event duration. The section of
volume $\Delta '$ then is a narrow crescent, of radius
$R_E(D)/A_{min}$ and width $V_{\perp}(t_{inter}+t_{exp})$, and~:
\beq
N_{events} = (t_{inter}+t_{exp}) N_{exp} \sum_{stars}
\left\{\int_{0}^{D_{star}} \frac{2 \, R_E(D)}{A_{min}} \,V_{\perp}
\,n(D,V_{\perp}) \,{\rm d} D {\rm d}^2V_{\perp} \right \}
\eeq
The product $(t_{inter}+t_{exp}) N_{exp}$ is the total observation
time $t_{obs}$ , and the expression between braces is nothing but
the event rate $\Gamma_{star}$, which becomes independent of the
star when the threshold amplification $A_{min}$ is constant and all
stars are nearly at the same distance.

\section {Proposed experiments}

Our initial idea to use the present setting of the Th\'emistocle
experiment (Kovacs 1990; Baillon 1992), in the south of France, proved
unrealistic. With only one photomultiplier counting photons from
the {\em whole} Andromeda galaxy M31, such an experiment could
detect a few events per year, but required a 10$^{-8}$ photometric
accuracy (and extreme stability). The field of view of a detector
must be small to detect with an achievable photometric accuracy the
lensing of one star. But the number of stars in the field of view
will then be small, and therefore the number of lensing events will
also be small. The solution is to add many detectors, or rather to
pixellize the detector. We studied two cases~: 10$^3$ pixels and
10$^6$ pixels. The first solution can be achieved with an array of Hybrid
PhotoDiode (HPD) tubes, and the second one with a typical CCD camera.

HPD tubes are equipped with a silicon wafer anode which converts
accelerated photoelectrons (20 keV or more) into 10$^4$ to 10$^5$
electron-hole pairs. Such HPD  tubes are under development for LHC experiments
(DeSalvo et al. 1991). Their main advantage is that they count
incoming photoelectrons one by one without loss and noise.
Therefore, the light flux that reaches a definite pixel is measured
with a very high accuracy.
Another advantage is
that they continuously record photons as they arrive on the
detector. Therefore, it is possible to look back at the record, and
add the number of photons received in bins of variable duration to
fit best the actual duration of a lensing event, which can range
between a few minutes and a few weeks, and to reconstruct the
typical shape of the micro-lensing light curve. Figure 3a shows how
this optimization of the exposure time improves the sensitivity of
the experiment. The main drawback of photomultipliers is that the
number of pixels cannot be very large, a few thousands at most. A
lensing event then has to emerge above a large background, and it
requires a very good photometric accuracy to be detected, at least
of the order of 10$^{-4}$. On the other hand, the relatively large
angular size of large pixels (of the order of one arcmin) does not
require a mirror of extremely good optical quality. To be precise,
we do our estimates for a set-up corresponding to a reduced version
of a planned Th\'emistocle upgrade, namely a mirror with a
collecting area of 1 m$^2$, and 1000 pixels. This can be realized
with 50 HPD tubes with 20 pixels each, fed by optical
fibers from the focal plane of a mirror.
We assumed in our estimations that these 1000
pixels were set in a flattened elliptical pattern (with a
semi-major axis of 30 arcmin, and a semi-minor axis of 7 arcmin),
to follow the elongated shape of the two targets that we chose, the
LMC and M31. The full upgrade will have two telescopes, each with a
12.5 m$^2$ mirror and between 1000 and 2000 pixels at the focus.

CCD cameras have advantages and drawbacks opposite to
photomultipliers. They have a very high spatial resolution and it
is easy to have several millions pixels on a small surface at a
reasonable cost. But their photometric accuracy is worse than
photomultipliers, around 1\% only. As a reference CCD camera, we
took the characteristics of the ongoing ``Naines Brunes''
experiment (Moniez 1990; Ansari 1992)~: a camera with 2$\times$8 CCD
chips of 564$\times$410 pixels each, at the focus of a 40 cm
diameter mirror. The photon detection efficiency was taken in both
cases to be 20\% in the V (visible) band, taking into account the quantum
efficiency of the detector, the absorption through the atmosphere,
and the mirror reflexion losses (Allen 1973). Parameters for the
2 detectors are summarized in Table 1.

\begin{center}
\begin{tabular}{|l|r|r|}
\hline
Detector & Photon counter & CCD Camera \\ \hline \hline
Collecting area (m$^2$) & 1 & 0.125 \\ \hline
Field of view & Elliptical 60'$\times$14' & Rectangular
64'$\times$22' \\ \hline
Number of pixels & 1000 & 3.7$\times 10^6$ \\ \hline
Aperture per pixel (arcsec$^2$) & 2300 & 1.36 \\ \hline
\end{tabular}

{\bf Table 1. Main parameters of the detectors.}
\end{center}

We consider two possible targets~: the Large Magellanic Cloud (LMC)
and the Andromeda galaxy M31. Each one has advantages, and
drawbacks~:

\noindent i) The LMC can only be seen from the southern hemisphere,
whereas M31 is best seen from the northern hemisphere.

\noindent ii) M31 is much further away than the LMC, about 14
times, and therefore a detector sees $14^2$ times more stars for
the same aperture. This has two consequences. First, to be detected
in M31, a lensing usually requires a larger amplification (see
Figure 1) of a brighter star (see Figure 2). Such an event is then
rarer, but this is almost exactly compensated by the larger number
of target stars. This larger average number of stars in a pixel has
a second consequence~: statistical fluctuations in the number of
stars in a pixel will be smaller, and therefore less precision is
required in pointing the telescope, and in matching pixels from one
image to another.

\noindent iii) If the target galaxy also is embedded in a brown
dwarf halo, then additional micro-lensings occur due to these brown
dwarfs. The Magellanic Clouds are embedded in our galactic halo and
(probably) have no halo of their own. On the contrary, the rotation
curves of M31 show that it is surrounded by a dark halo. Therefore
brown dwarfs of this halo can act as gravitational lenses for stars
in M31, and we must add their contribution. To be conservative, we
assumed the same halo for M31 as for our galaxy, although the
larger rotation velocities imply a larger dark matter density
around M31. We find that the contribution from the brown dwarfs of
M31 is slightly larger than the contribution from our own galaxy
for heavier brown dwarfs (i.e. with masses in the range $10^{-3}$
to $10^{-1} \,M_{\odot}$) but negligible for lighter masses (see
Figure 3b).

\section{Results}

The foremost result is that the expected number of micro-lensing
events is about one order of magnitude larger in both our proposals
than in ongoing experiments (Alcock et al. 1989; Griest 1991; Ansari 1992),
 even
though neither detectors were optimized for this brown dwarf
search. We call an event ``detected'' if the luminosity of a
definite pixel stays at 3 standard deviations above the background
(that is $Q = 3$ in Equation \ref{amin2}) for three consecutive
exposures, and rises above 5 during one of them. In fact we often
``found'' events at $Q > 10$.

Let us first focus on the photomultiplier array, pointing to M31.
Figure 3a shows the number of detected events as a function of the
brown dwarf mass $M_{bd}$, for a duty cycle of 120 nights of 6
hours, and for three different exposure times. In  such experiments the best
sensitivity is obtained for brown dwarfs
in the mass range 10$^{-5}$ to 10$^{-3}M_{\odot}$. The brown dwarf
mass density being fixed, heavier brown dwarfs are less numerous
and produce less lensing events. On the other hand, very light brown dwarfs,
although more numerous, are not
very efficient because their maximal amplification is low, and
moreover they produce short lensing events, which can be too short
to be detected.  In this respect, short exposures (1/4
hour) are adequate to detect light brown dwarfs (10$^{-7}$ to
10$^{-4}$ M$_{\odot}$), whereas high mass brown dwarfs (10$^{-3}$
to 10$^{-1}$ M$_{\odot}$) generate long lensing events and are best
detected with exposures of the order of 6 hours. These results show
the interest of the photomultiplier device, which allows the tuning
of the exposure time to the mass of the brown dwarf~: incoming
photons can be recorded one by one with a very short time binning,
the time bins can then be lumped together at will.

Figure 3b shows the relative importance of brown dwarfs from our
own halo and of brown dwarfs from the halo of M31. As we said, both
halos equally contributes above $M_{bd}=10^{-5}M_{\odot}$, but the
importance of M31 halo vanishes for lighter masses. This is due to
the finite size effect~: Equation \ref{amax} shows that for
amplifications $A \gg 1$~:  \beq
A_{max} \simeq 2 \frac{R_E(D) / D}{R_{star}/D_{star}}
\eeq
that is, twice the ratio of the angular diameter of the Einstein
ring to the angular diameter of the star. The Einstein radius $R_E$
is about the same for brown dwarfs in our halo and in M31 halo, but
the {\em angular} diameter is much smaller for M31 brown dwarfs,
and the maximal amplification is therefore much smaller.

The exposure time cannot easily be varied for CCD's, and in any
case cannot be changed after the exposure. CCD's must be read after
a definite time, long enough to get a signal sufficiently above the
reading noise, but short enough to avoid saturation. The exposure
time was taken to be 1/4 h. Longer exposures can be achieved by
combining successive images, but this is very expensive in
computing time due to the large number of pixels, and the gain in
statistical accuracy can be lost due to the difficulties in
matching pixels between the combined images. For this reason, in
the case of CCD cameras, we only show in Figure 4 the expected
counting rates for exposures of 1/4 h. As expected, the plot is
strongly peaked towards low masses when looking at the LMC, just
because the number density of brown dwarfs is larger for lighter
masses. The decrease at the lower end of the mass range again
reflects the shorter duration of lensing events. The decrease is
stronger when the target is M31, again because of the finite size
effect~: Figure 2 tells us that target stars in M31 are very
luminous, and most of them are red giants with large diameters that
limit the possible amplifications.

To check the sensitivity of our results to the luminosity function
of the target galaxy, we compare in Figure 4 the counting rates obtained
taking for the bright stars of M31 either the true data (Hodge et al. 1988)
or the solar neighborhood data (Allen 1973). Clearly the change is
not significant, and our poor knowledge of the luminosity function and
spectral repartition does not affect the results.

How do our result compare with ongoing brown dwarf searches~? Table
2 is a summary of our results~: we compare our expected counting
rates for different target galaxies (LMC {\em vs} M31) and
different detectors (CCD camera {\em vs} HPD tubes), to
those announced by the ``Naines Brunes'' CCD experiment
(Ansari 1992). However our evaluations do not take into account
all experimental cuts.

\begin{center}
{\small
\begin{tabular}{|l|r|r|r|r|r|}
\hline
Experiment & Naines Brunes & CCD$\rightarrow$LMC &
CCD$\rightarrow$M31 & PM$\rightarrow$LMC & PM$\rightarrow$M31 \\
\hline \hline
$N_{events}$ & 10 & 80 & 50 & 60 & 60 \\ \hline
$<A_{min}>$ & 1.34 & 9.6 & 23 & 7 & 14 \\ \hline
$<{\rm Mag}_V>$ & $<\,0$ & 2 & -1 & 1 & -2 \\ \hline
Photometric & 10$^{-2}$ & 10$^{-2}$ & 10$^{-2}$ & 10$^{-4}$ &
10$^{-4}$ \\
accuracy & & & & & \\ \hline
Optimal $M_{bd}$ & $10^{-6}$ & $10^{-6}-10^{-5}$ & $10^{-4}$ &
$10^{-5}$ & $10^{-4}$ \\ \hline
Lensed stars & Red giants & Main sequence & Red giants & Main
sequence & Red giants \\ \hline
\end{tabular}
}

{\bf Table 2. Main results of simulations.}
\end{center}

What does Table 2 tells us~? First, the expected number
$N_{events}$ of detected micro-lensings is larger when pixels are
monitored instead of stars. The reason is that the ``Naines
Brunes'' experiment only monitors stars of absolute magnitude less
than 0 in the LMC, and is not sensitive to fainter stars. Figure 2
shows the distribution of the absolute magnitude of the lensed
stars in the 'detected' events in our CCD$\rightarrow$LMC
Monte-Carlo simulation. The shaded histogram corresponds to the
LMC, and one can see that stars fainter than magnitude 0 represent
nearly 90\% of our detected events, hence the large increase in
sensitivity from 10 events/year to about 80.

The expected number of events accidentally turns out to be similar
when the target is M31 or the LMC, or with a photomultiplier array
instead of a CCD camera. Of course, events are different in each
experiment.The requested amplifications $A_{min}$ are of course
larger than the value 1.34 of the ``Naines Brunes'' experiment. The
requirement of higher amplifications for our proposals implies that
they are automatically sensitive to higher brown dwarf masses than
ongoing experiments. Amplifications are smaller for the
photomultiplier experiment than for the CCD experiment, only
because the photometric precision is much better and compensates
the larger background due to the much larger number of stars on a
pixel. The photometric accuracy 1/$\sigma$ due to statistical
fluctuations of the photon flux,
computed from Equation \ref{sigma}, is 10$^{-2}$ for the CCD
camera and
10$^{-4}$ for the HPD array. It can be much smaller
with a larger collecting area (for instance going from the 1 m$^2$
of our HPD example to the 25 m$^2$ of one Th\'emistocle upgrade), but
other sources of noise may then dominate.

\section{Going further~?}

The estimates above were not performed for optimized
experiments, but for existing (or planned) detectors built
for other purposes. However, before a dedicated experiment can be
undertaken, some proof of concept is obviously needed. Many weak
points were left over in this work. One of the most blatant is
the neglect of spurious fluctuations due to variables stars and
pointing errors or bad pixel matching. Variable stars can in
principle be distinguished from micro-lensing events, because in
the latter case the light curve must have only one maximum, be time
symmetric and achromatic. All these
characteristics also hold in our case, but may be more difficult to
check because we rely on a small flux increase over a large
background. The achromaticity of the light {\em increase}
 could be checked by taking pictures in
two colours, provided that the signal is strong enough in {\em
both} colours.
Pointing errors have less serious
consequences for a distant target such as M31 than for the LMC,
because luminosity fluctuations are much smaller from one pixel to
the next. The large pixels of photomultipliers also lead to smaller
fluctuations, but as a higher photometric precision is needed,
the situation is not better than for CCD cameras. These
fluctuations can, to a large extent, be corrected if one requires
that bright spots be at the same position on every picture.

All these problems are now under study, but a rudimentary test
of our proposal could be done, for instance, using data already
gathered by
the ``Naines Brunes''  experiment (Ansari 1992). This experiment
first
produces a star catalog from a few CCD pictures, using pattern
recognition.
To decrease photometric errors and computer time,
the remaining CCD pictures are then processed to extract the
luminosity curves of the stars already in the catalog. A faint
star absent from the catalog will not be monitored, but may appear
on a few pictures when lensed by a brown dwarf. Such lensing events
could be detected as transient stars, if the pattern recognition
algorithm
could be applied to all CCD pictures.
A preliminary study indicates that the number of detected
events could be doubled that way, but some work is needed to
evaluate the background of fake events.\\

\noindent {\Large \bf References}

\noindent Alcock C., Axelrod T., and Park H.~S., October
1989, Seminar at Center for Particle Astrophysics Berkeley.

\noindent Allen C.~W., 1973, { Astrophysical Quantities},
Athlone Press, London.

\noindent Ansari R., 1992, IVth Rencontre de Blois, Blois
, Fontaine G. et al., (eds.), Editions Fronti\`eres, Gif sur
Yvette, France.

\noindent Ardeberg A. et~al., 1985, {A\&A} 148, 263.

\noindent Bahcall J.~N. and Soneira R.~M., 1980, {ApJS} 44, 73.

\noindent Baillon P. et al., 1992, Detection of very high energy
Gamma rays from the Crab source, in: Proceedings of I.C.H.E.P., Dallas.

\noindent Caldwell J.~{A. R.} and Ostriker J., 1981, {ApJ} 251, 61.

\noindent Carr B.~J., Bond J.~R. and Arnett W.~D., 1984, {ApJ} 277, 445.

\noindent De R\'ujula A., Jetzer Ph. and Mass\'o \'E., 1992,
{ A\&A} 254, 99.

\noindent DeSalvo R., Hao W., You K., Wang Y. and Xu C., 1992, Nucl. Instr. \&
Meth. A315, 375.

\noindent de Vaucouleurs G. and Freeman K. C., 1972, { Vistas
in Astronomy} 14,163.

\noindent Faber S.~M. and Gallagher J.~S., 1979, { ARA\&A} 17, 135.

\noindent Flores R.~A., 1988, { Phys. Lett. B} 215, 73.

\noindent Griest K., 1991, { ApJ} 366, 412.

\noindent Hardy E. et~al., 1984, { ApJ} 278, 592.

\noindent Hodge P. et~al, 1988, { ApJ} 324, 172.

\noindent Kent S.M., 1983, {ApJ} 266, 562.

\noindent Kormandy J. and Knapp G.~R. (eds.), 1987, Proc. IAU Symp. 117,
Dark matter in the Universe. Reidel.

\noindent Kovacs F., 1990, { Nucl. Phys. B (Proc. Supp.)} 14A, 330.

\noindent Moniez M., 1990, in: Blanchard A. et al., (eds.) Xth Moriond
astrophysics meeting. Editions Fronti\`eres,
Gif sur Yvette, France.

\noindent Paczy\'nski B., 1986., { ApJ} 304, 1.

\noindent Rees M.~J., 1986, in: Baryonic dark matter, 2nd
ESO-CERN Symposium .

\noindent Trimble V., 1987, { ARA\&A}, 25, 425.

\noindent Turner M.~S., 1990, Dark matter in the universe,
Technical Report Fermilab-Conf 90/230A, FERMILAB. Talk at
Nobel Symposium 79 'The birth and early evolution of our universe'
(Graftavallen73, 11-16 June 1990).

\noindent Walterbos R. A. M. and Kennicutt R. C. Jr., 1987., {
A\&AS} 69, 311.\\

\noindent {\Large {\bf Figure captions}}

\noindent {\bf Fig. 1.} Amplification distribution for the CCD
camera pointing at the LMC or M31, for a brown dwarf mass
$M_{bd}=10^{-4}M_{\odot}$ and an exposure time $t_{exp}$ = 1/4 h.

\noindent {\bf Fig. 2.} Magnitude distribution of the lensed
stars for the CCD camera pointing at the LMC or M31, for a brown
dwarf mass $M_{bd}=10^{-4}M_{\odot}$ and an exposure time $t_{exp}$
= 1/4 h.

\noindent {\bf Fig. 3. a} Number of detected micro-lensing events
per year for the photomultiplier array pointing to M31, as a
function of the brown dwarf mass. The curves correspond to 3
different exposure times, $t_{exp}$ = 1/4~h, 1~h and 6~h.
\newline {\bf b} Number of detected micro-lensing events
per year for the photomultiplier array pointing to M31, as a
function of the brown dwarf mass. The exposure time was optimized
according to the brown dwarf mass. The curves show the number of
events due to brown dwarfs of our halo, of the halo of M31, and the
total number of events.

\noindent {\bf Fig. 4.} Number of detected micro-lensing events
per year for the CCD camera pointing to the LMC and to M31, as a
function of the brown dwarf mass. The solid curves are obtained with the
true luminosity function of the LMC (Hardy et al. 1984) and of M31
(Hodge etal. 1988). The dashed curve uses the solar neighborhood data
over the whole range of magnitudes.

\end{document}